\documentclass[11pt]{article}
\usepackage{graphicx}  
\usepackage{moriond,epsfig}

\def\Journal#1#2#3#4{{#1} {\bf #2}, #3 (#4)}

\def\NIMA{{\em Nucl. Instrum. Methods} A}

\def\PLB{{\em Phys. Lett.}  B}
\def\PRL{\em Phys. Rev. Lett.}
\def\PRD{{\em Phys. Rev.} D}

\def\be{\begin{equation}}
\def\ee{\end{equation}}
\def\bea{\begin{eqnarray}}
\def\eea{\end{eqnarray}}

\begin{document}
\vspace*{4cm}
\title{$D$ Hadronic Analyses at CLEO}

\author{ {\it The CLEO Collaboration} \\ M.S. DUBROVIN }

\address{Department of Physics \& Astronomy, Wayne State University,\\
Detroit, MI 48201}

\maketitle
\abstracts{
The CLEO-c results on $D$ meson production and hadronic decays
obtained with currently available data sets are presented.}

%%%==================================================
\section{Introduction}
Recent CLEO-c results on $D$ meson production and hadronic decays are presented in this overview.
The CLEO-c is a general purpose detector at CESR, Cornell Electron-positron Storage Ring.
The detector configuration is an upgraded version of the CLEO III~\cite{CLEOIII}.
In order to run at charm production energy the Silicon Vertex Detector was replaced by the 
6 stereo-layers drift chamber; magnetic field in the superconducting solenoid is reduced to 1~T
for optimal momentum resolution at lower energy. For stable operation at low energy  
the 12 superconducting wigglers have been installed at CESR upgrade.

The CLEO-c experimental program~\cite{YellowBook} was started in October 2003 and
will continue until April 2008.
Analyses included in this overview use three samples of events.
\\ 1) 
At 3770~MeV we have collected luminosity $\sim$560~pb$^{-1}$, that corresponds to about 
4~M produced $\psi(3770)$ dominantly decaying to $D\bar{D}$ pairs.
The 281~pb$^{-1}$ of this sample are processed. We plan to accumulate more data
at $\psi(3770)$ by the end of runs. 
\\ 2)
The 12 points scan in the range from 3970 to 4260~MeV 
with total luminosity of $\sim$60~pb$^{-1}$ is performed
in order to find an optimal energy for $D_s$ meson study.
We find that the $D^\pm_s D_s^{*\mp}$ cross section is maximal around 4170~MeV.
\\ 3)
At 4170~MeV we have collected 314~pb$^{-1}$, of which 195~pb$^{-1}$ are processed.
At this energy we plan to accumulate 750~pb$^{-1}$ in total. 

Below I discuss four topics:
{\it (i)} 
the absolute $D^0$, $D^+$, and $D_s$ meson hadronic branching fractions measurement;
{\it (ii)} 
the $e^+e^- \to D_{(s)}^{(*)} \overline{D}_{(s)}^{(*)}  $ cross section measurement 
in the energy range from 3970 to 4260~MeV;
{\it (iii)} 
the inclusive $\eta$, $\eta '$, and $\phi$ meson production branching fractions measurement 
in the $D^0$, $D^+$, and $D_s$ decays; and
{\it (iv)} 
the precision $D^0$ mass measurement.  
Non-covered recently published results and ongoing analyses are also listed below.

%%%==================================================
\section{Absolute hadronic branching fractions of $D^0$ and $D^+$ mesons}
\label{sec:HadBF}
\begin{figure}[t]
  \begin{minipage}[t]{50mm}
  \includegraphics[width=50mm]{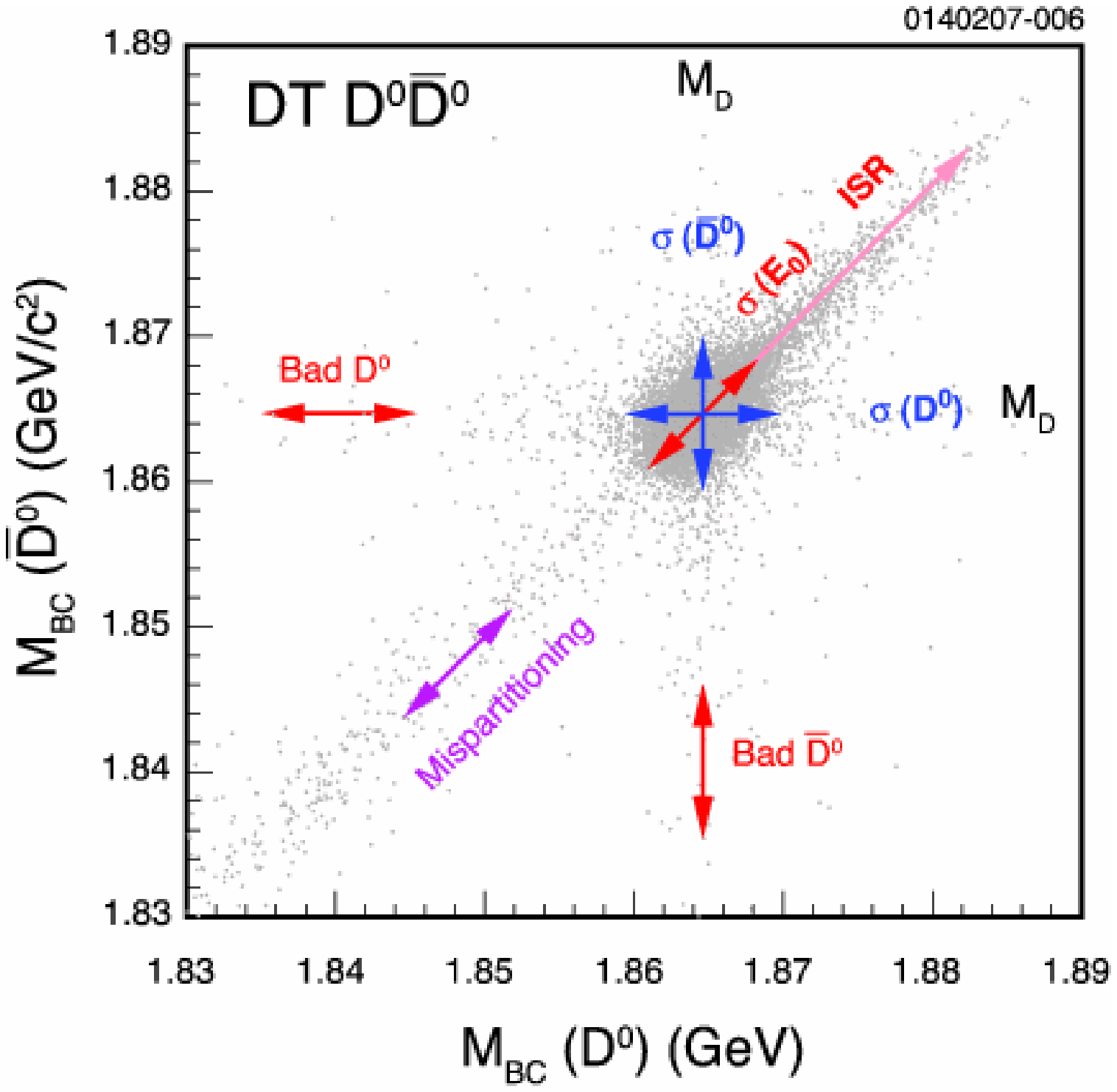} 
  \caption{\label{fig:DT_2D} The $m_{BC}(\bar{D})$ {\it vs.} $m_{BC}(D)$  distribution in MC.}
  \end{minipage}
\hfill
  \begin{minipage}[t]{100mm}
  \includegraphics[width=98mm]{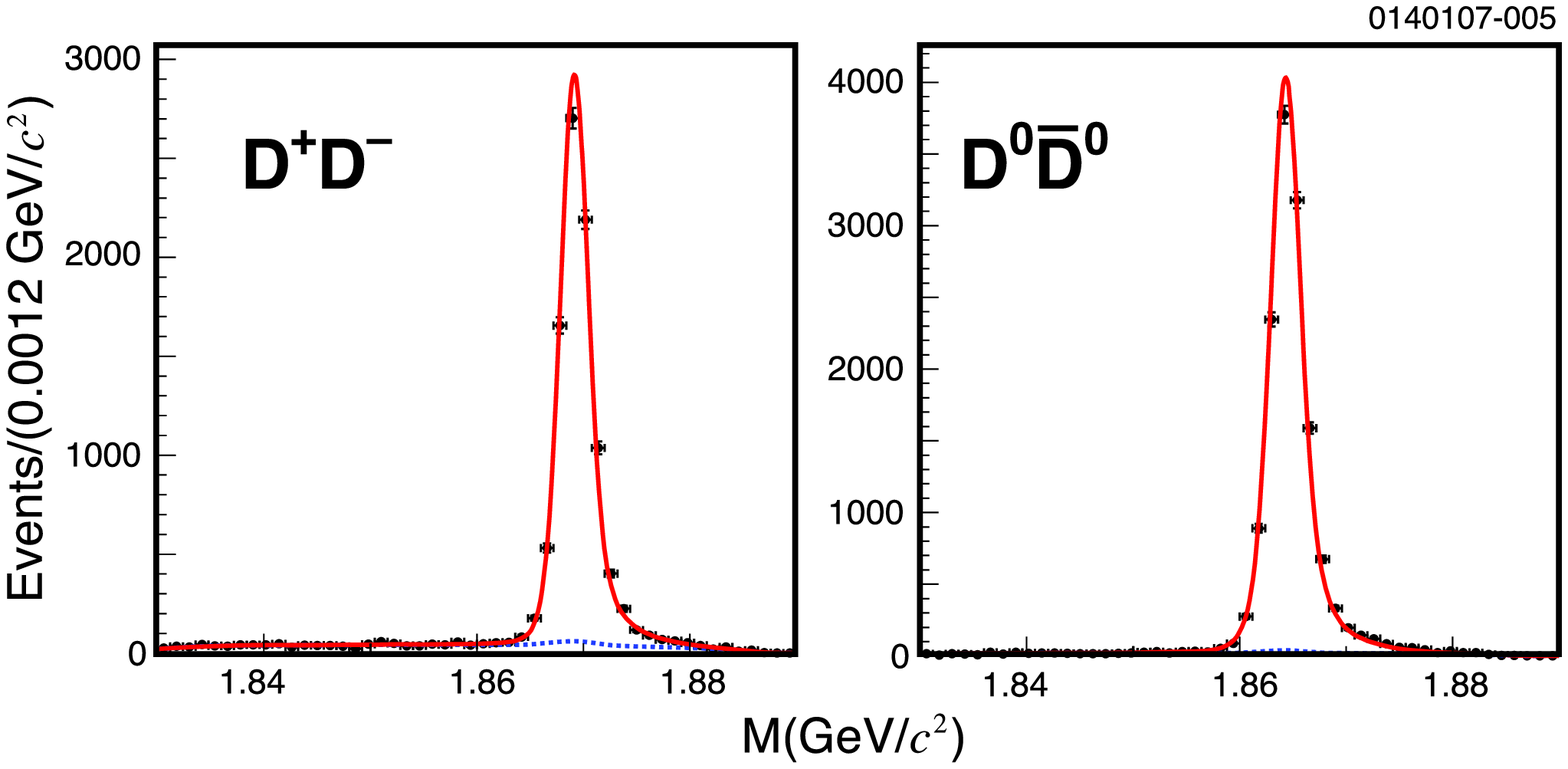}
  \caption{\label{fig:DT_1D} The $m_{BC}$ distributions for DT events combined for all $D^0$ and $D^+$ modes.}
  \end{minipage}
\end{figure}

\begin{figure}[t]
  \includegraphics[width=140mm]{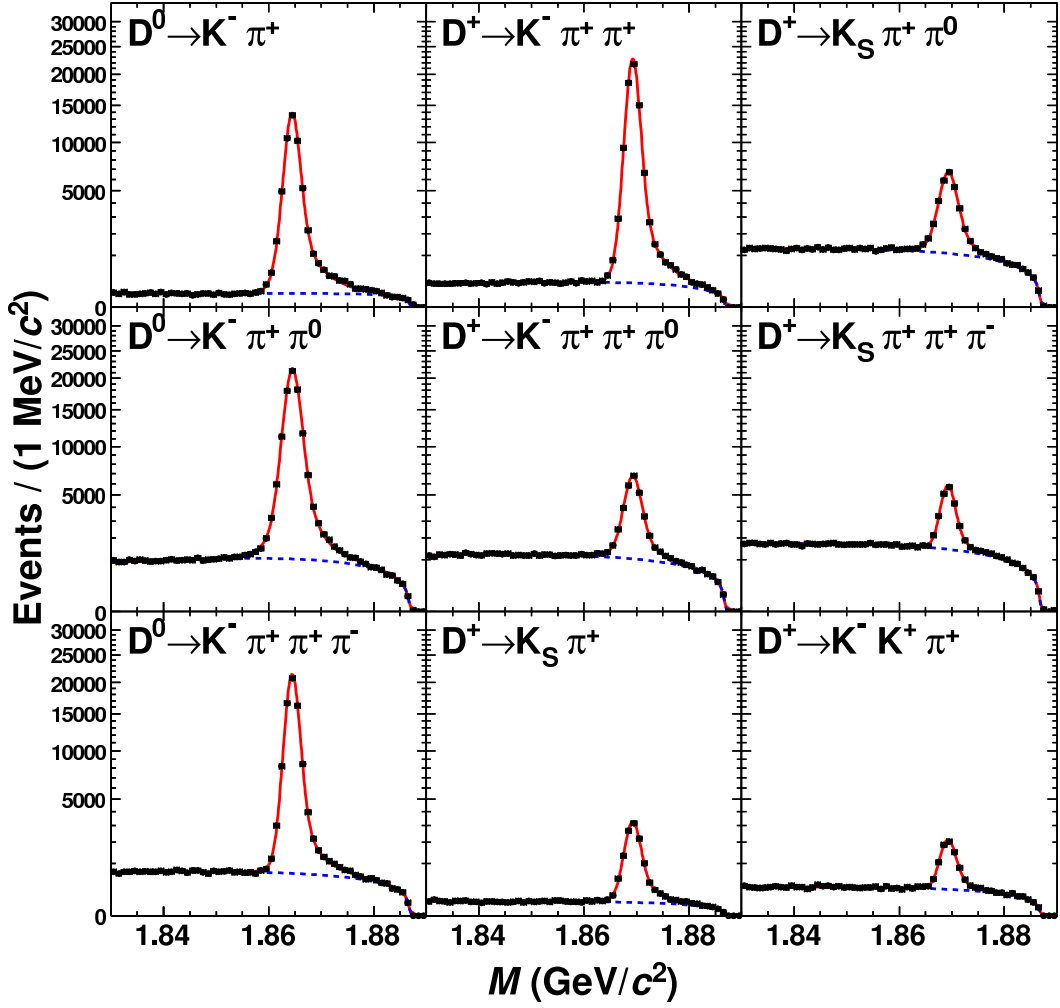}
  \caption{\label{fig:ST_combined} The $m_{BC}$ distributions for three $D^0$- 
                                   and six $D^+$-meson decay modes.}
\end{figure}
%
%%%-------------------------------
%
\begin{table}[b]
\begin{minipage}[t]{75mm}
\caption{\label{tab:HadBF} Hadronic branching fractions of $D^0$ and $D^+$. }
%\scriptsize
\begin{center}
\begin{tabular}{|l|c|}
\hline
%\multicolumn{9}{|c|}{ txt } \\
Mode & $\cal B$, \% \\
\hline
$D^0 \to K^-\pi^+$              &   3.87$\pm$0.04$\pm$0.08 \\
$D^0 \to K^-\pi^+\pi^0$         &   14.6$\pm$0.1$\pm$0.4   \\
$D^0 \to K^-\pi^+\pi^+\pi^-$    &   8.3$\pm$0.1$\pm$0.3    \\
\hline
$D^+ \to K^-\pi^+\pi^+$         &   9.2$\pm$0.1$\pm$0.2    \\
$D^+ \to K^-\pi^+\pi^+\pi^0$    &   6.0$\pm$0.1$\pm$0.2    \\
$D^+ \to K^0_S\pi^+$            &   1.55$\pm$0.02$\pm$0.05 \\
$D^+ \to K^0_S\pi^+\pi^0$       &   7.2$\pm$0.1$\pm$0.3    \\
$D^+ \to K^0_S\pi^+\pi^+\pi^-$  &   3.13$\pm$0.05$\pm$0.14 \\
$D^+ \to K^+K^-\pi^+$           &   0.93$\pm$0.02$\pm$0.03 \\
\hline
\end{tabular}
\end{center}
%\end{table}
\end{minipage}
\hfill
\begin{minipage}[t]{75mm}
%\begin{table}[!htb]
\caption{\label{tab:Ds_HadBF} The $D_s^+$ hadronic branching fractions.}
%\scriptsize
\begin{center}
\begin{tabular}{|l|c|}
\hline
%\multicolumn{9}{|c|}{ txt } \\
Mode                           &  $\cal B$, \%          \\
\hline
$D_s^+ \to K^0_S K^+$          & 1.50$\pm$0.09$\pm$0.05 \\
$D_s^+ \to K^-K^+\pi^+$        & 5.57$\pm$0.30$\pm$0.19 \\
$D_s^+ \to K^-K^+\pi^+\pi^0$   & 5.62$\pm$0.33$\pm$0.51 \\
$D_s^+ \to \pi^+\pi^+\pi^-$    & 1.12$\pm$0.08$\pm$0.05 \\
$D_s^+ \to \pi^+\eta$          & 1.47$\pm$0.12$\pm$0.14 \\
$D_s^+ \to \pi^+\eta '$        & 4.02$\pm$0.27$\pm$0.30 \\
\hline
\end{tabular}
\end{center}
\end{minipage}
\end{table}
%%%-------------------------------
%
The measurement of absolute hadronic branching fractions of $D^0$ and $D^+$ meson decays
is early published \cite{DHadBF} for luminosity 56~pb$^{-1}$. 
Now it is updated for 281~pb$^{-1}$.
We use technique pioneered by MARK III collaboration:
count the yield of the single tags (ST), events where the one side $D$ of $D\bar{D}$ pair
is reconstructed in one of considered hadronic modes, the other side $D$ is ignored;
count the yield of the double tags (DT), events where both $D$ mesons are reconstructed 
in considered hadronic modes. 
Efficiency is defined from full MC simulation. 
The total number of $D\bar{D}$ pairs 
produced and branching fractions are extracted from the $\chi^2$ fit.
The yields are defined using the signal variables 
$\Delta E = E_D - E_{\rm{beam}}$ and 
$m_{BC} = \sqrt{E^2_{\rm{beam}} - P^2_D}$, 
where $E_{\rm{beam}}$ is a beam energy, $E_D$ and $P_D$ 
are the reconstructed $D$ meson candidate energy and momentum, respectively.
Typical resolutions are 
$\sigma(\Delta E) = 7-10$~MeV and 
$\sigma(m_{BC}) = 1.3$~MeV/$c^2$ for modes comprising of tracks only.
Presence of $\pi^0$ meson in the final state degrades resolution by a factor of two roughly.
The DT yields are extracted from 2D fit to the 
$m_{BC}(D)$ {\it versus} $m_{BC}(\bar{D})$ distributions.
One of them is shown in Fig.~\ref{fig:DT_2D} for MC events. 
Fit accounts for $m_{BC}$ resolutions, beam energy spread, ISR,
mis-reconstruction of one-side or both D mesons.
Projections of the 2D scatter-plots and relevant yields for $D^0\bar{D^0}$ 
and $D^+D^-$ events are shown in Fig.~\ref{fig:DT_1D}.
The ST yields are extracted from fit to the $m_{BC}$ distributions, Fig.~\ref{fig:ST_combined}.
We have considered three decay modes for $D^0$ and six for $D^+$ meson 
with largest branching fractions.
Fit uses ARGUS function for the background shape and ``first principles'' in order to 
parameterize the signal component: the $m_{BC}$ resolution, ISR, Breit-Wigner shape
for $\psi(3770)$. The total number of reconstructed ST is $\sim$230K for $D^0$,
and $\sim$167~K for $D^+$ meson.
Results for nine branching fractions are shown in Table~\ref{tab:HadBF}.
We compare them with PDG-2004~\cite{PDG-2004} because PDG-2006~\cite{PDG-2006} 
includes our results from 56~pb$^{-1}$ analysis \cite{DHadBF}. 
The six of measured branching fractions are consistent with the world average
values~\cite{PDG-2004} within one standard deviation, 
the three -- within two standard deviations. Little has changed from 56~pb$^{-1}$ sample, but
systematic uncertainties dominate now. It should be noted that the 
final state radiation (FSR) effect is included in the efficiency calculation. 
Without account of the FSR the branching fractions decrease up to 2\% (depending on mode).

%%%==================================================
\section{$D$ meson pair production cross sections}
\begin{figure}[t]
  \begin{minipage}[t]{100mm}
  \includegraphics[width=50mm]{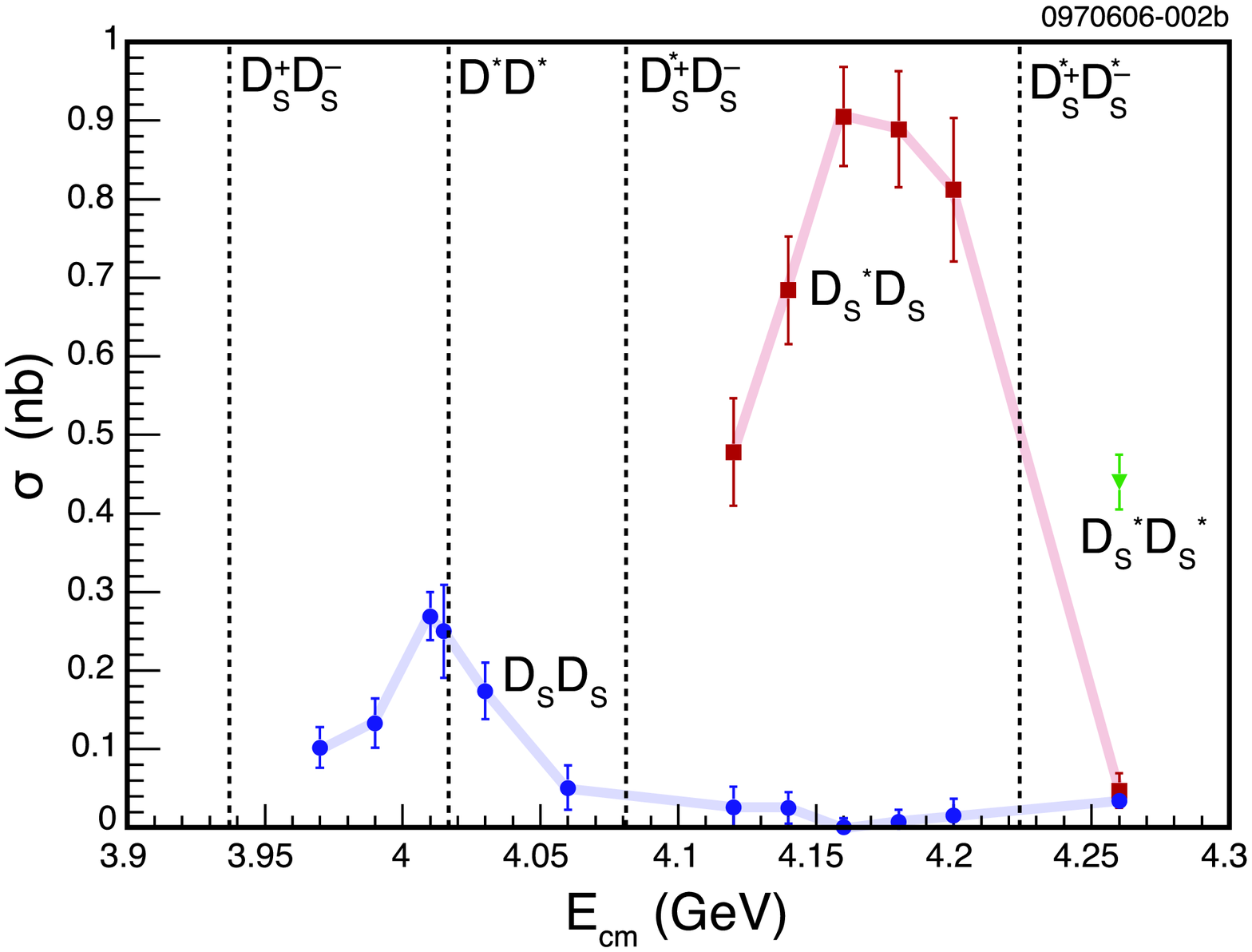}
  \includegraphics[width=49mm]{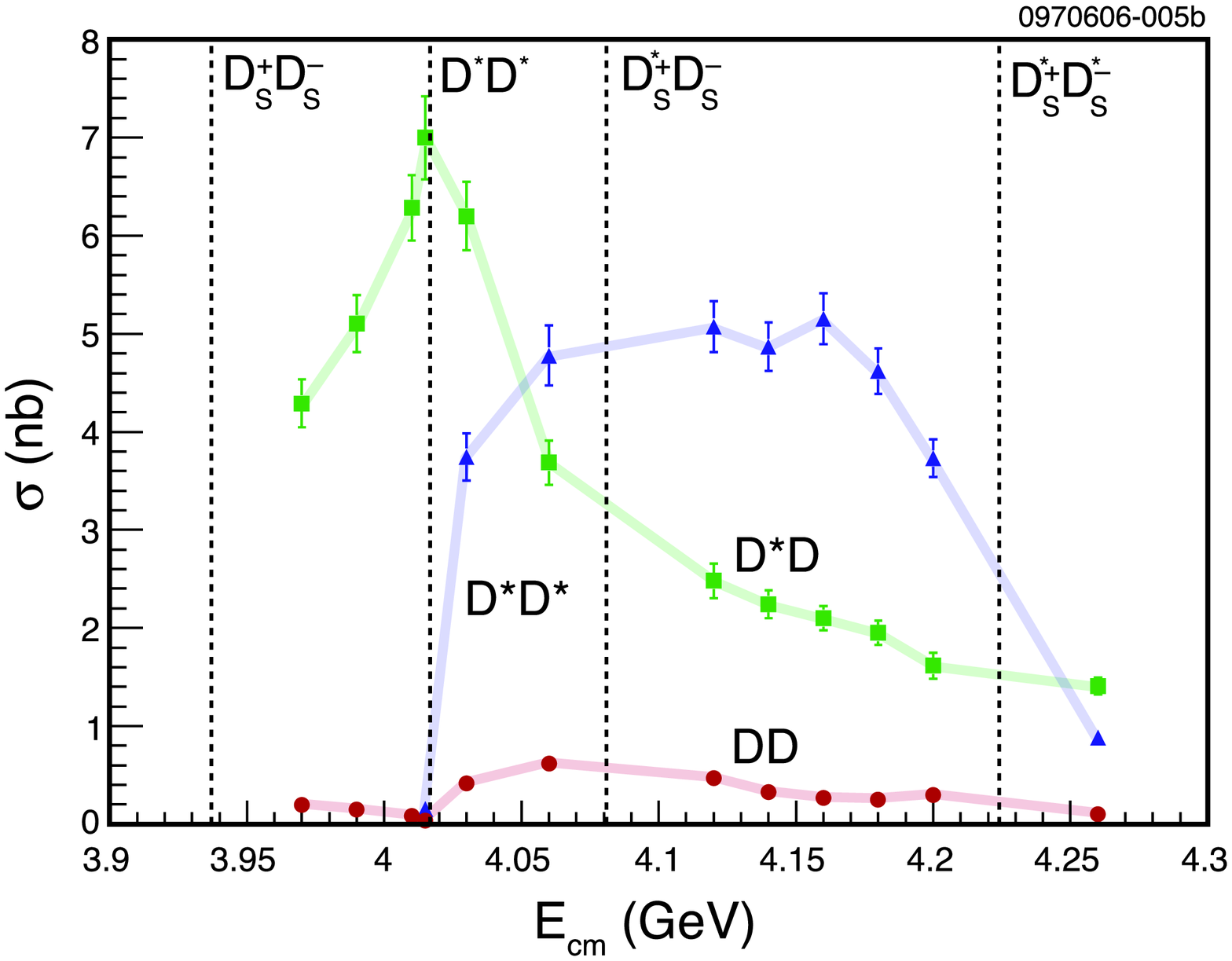}
  \caption{\label{fig:DDbar_cross-sections} The production cross sections
                                            for different $D$-meson pair combinations.}
  \end{minipage}
  \hfill
  \begin{minipage}[t]{56mm}
  \includegraphics[width=56mm]{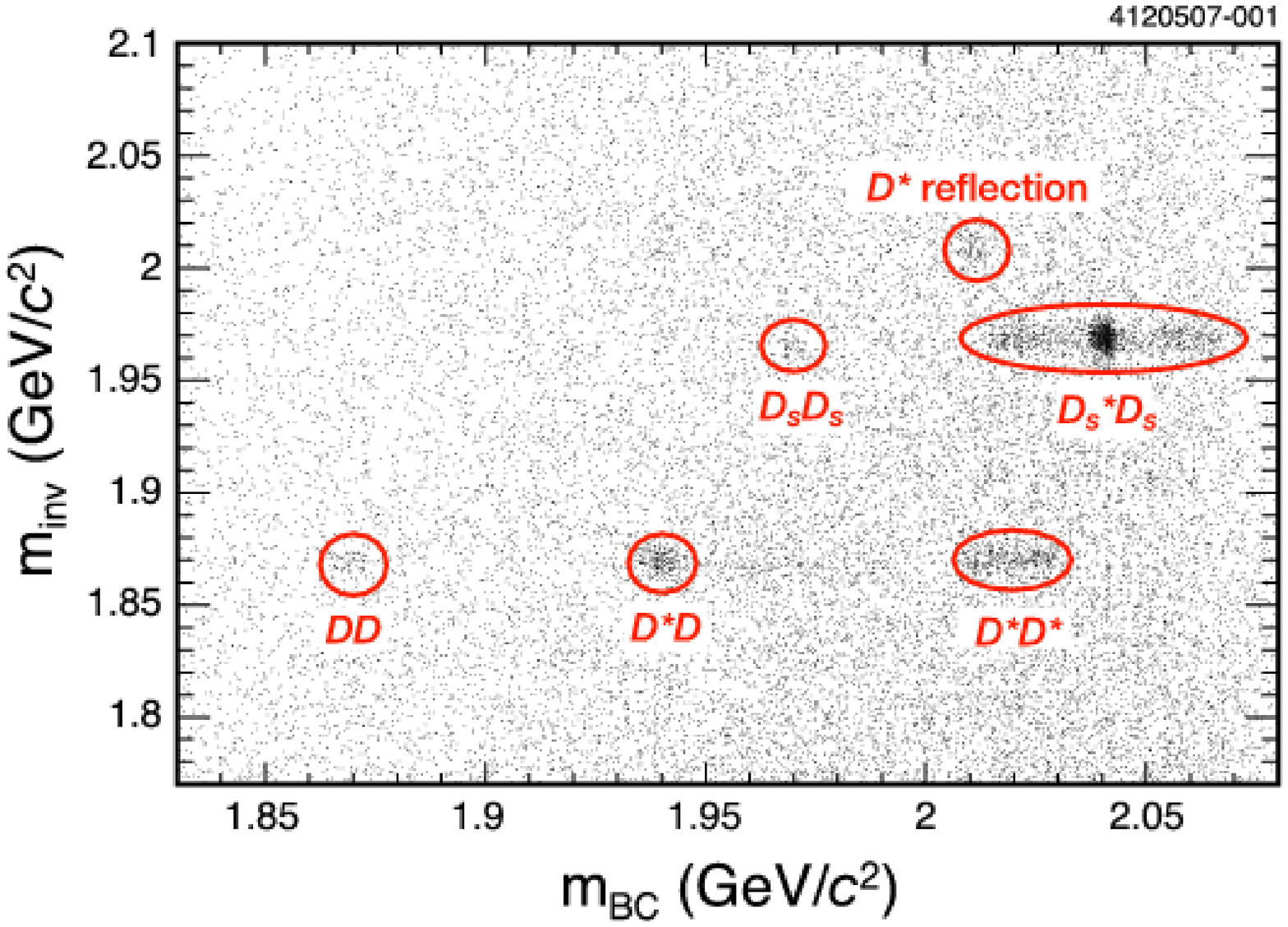}
  \caption{\label{fig:DDmodes_separation} Kinematic separation of the different 
                                          $D$-meson pair combinations.}
  \end{minipage}
\end{figure}
In order to find an optimal energy for $D_s$ meson study we 
performed the energy scan with total luminosity of 60~pb$^{-1}$ in 12 points from 3970 to 4260~MeV.
We launched this study because in earlier experiments the
total hadronic cross section was only measured.
We have measured six pair production cross sections,
$D\bar{D}$, $D\bar{D^*}$, $D^*\bar{D}^*$ , $D_s^\pm D_s^\mp$ , $D_s^{\pm}D_s^{*\mp}$, 
and $D_s^{*\pm}D_s^{*\mp}$, shown in Fig.~\ref{fig:DDbar_cross-sections}.
The largest $D_s$ meson production rate can be achieved at energy around 4170~MeV in the process
$e^+e^- \to D_s^{\pm}D_s^{*\mp}$, which cross section reaches 0.9~nb. 

%%%==================================================
\section{Absolute $D_s$ hadronic branching fractions}
\begin{figure}[t]
  \begin{minipage}[t]{104mm}
  \includegraphics[width=58mm]{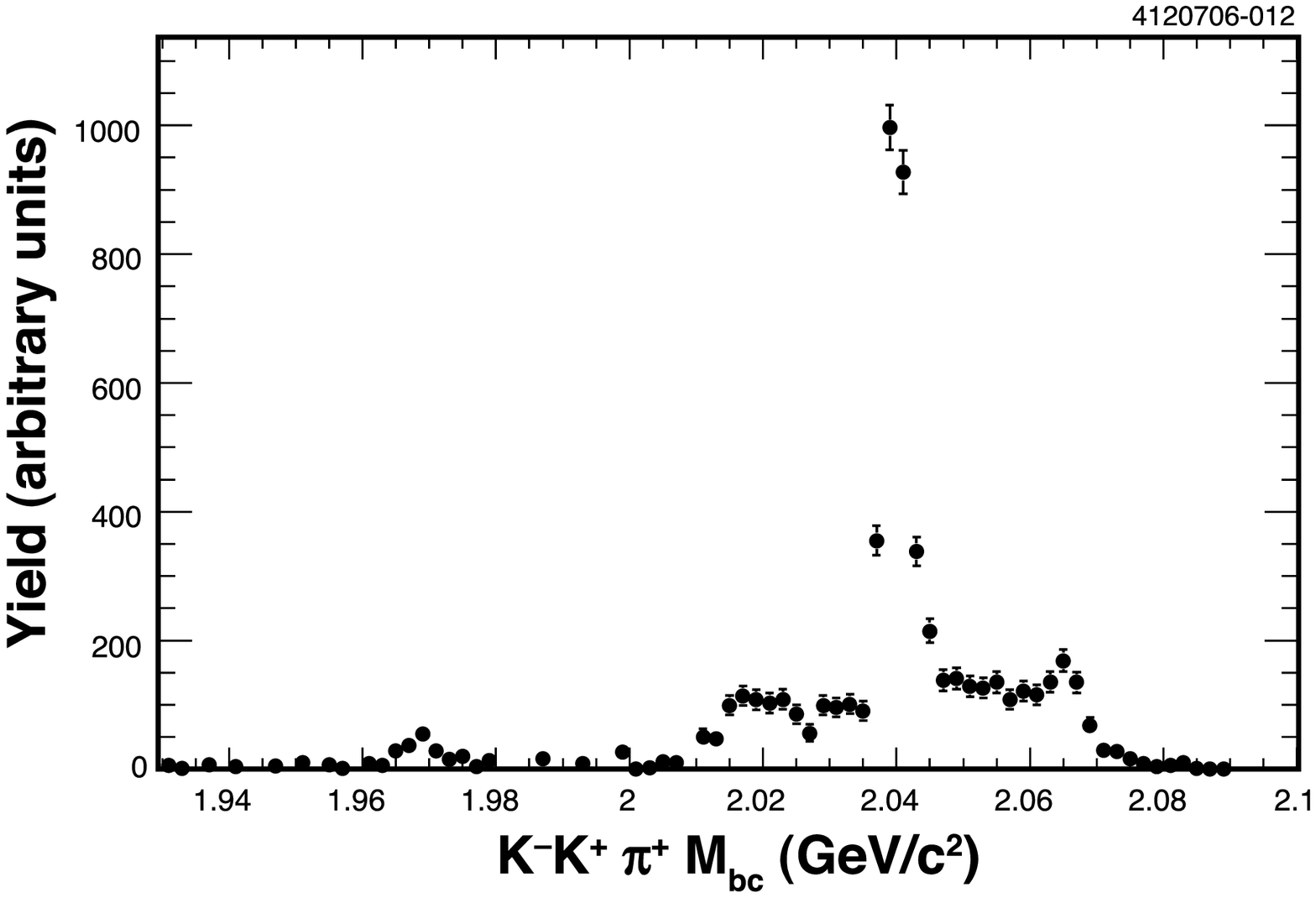} \hfill
  \includegraphics[width=45mm]{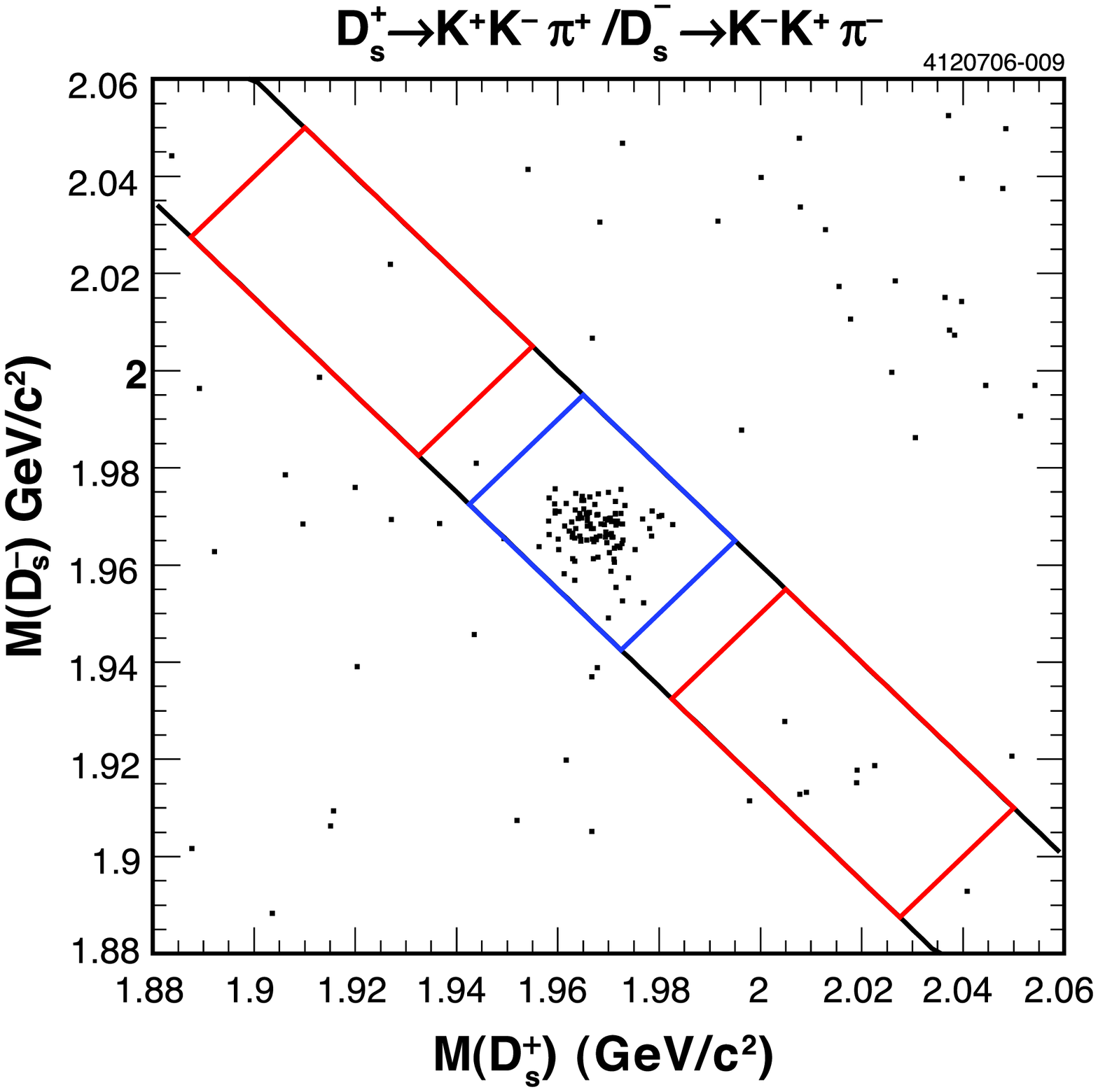} 
  \caption{\label{fig:Ds_HadBF_mass_spectra} The $m_{BC}(K^-K^+\pi^+)$ distribution (left), and
                                             the 2D invariant mass scatter-plot 
                                             for DT candidates (right). 
                                             The central box and two sideband boxes show the signal and
                                             background regions, respectively.}
  \end{minipage}
  \hfill
  \begin{minipage}[t]{54mm}
  \includegraphics[width=54mm]{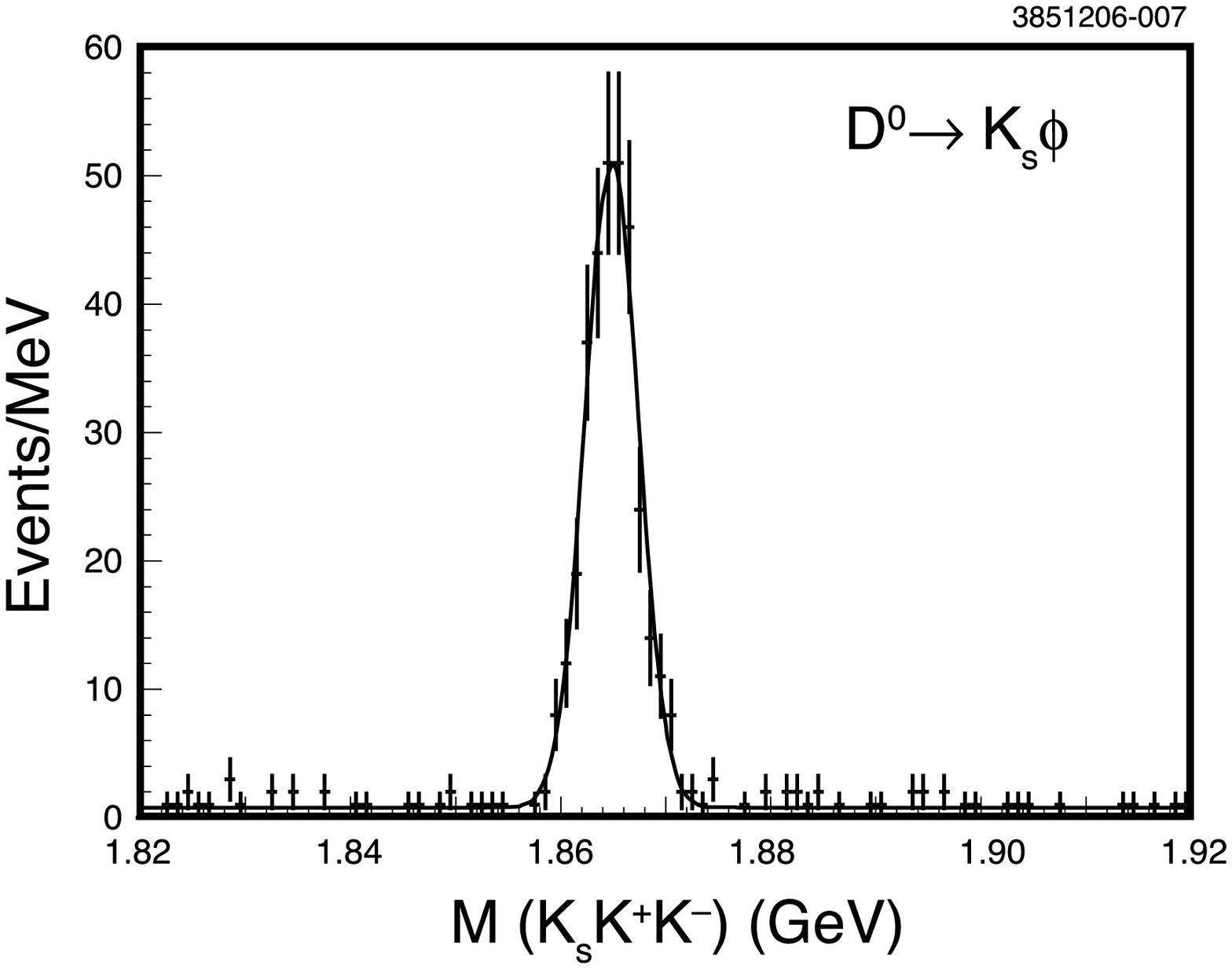}
  \caption{\label{fig:D0_mass_spectrum} The $K^0_SK^+K^-$ invariant mass spectrum.}
  \end{minipage}
\end{figure}
\begin{figure}[t]
  \hfill
  \includegraphics[width=50mm]{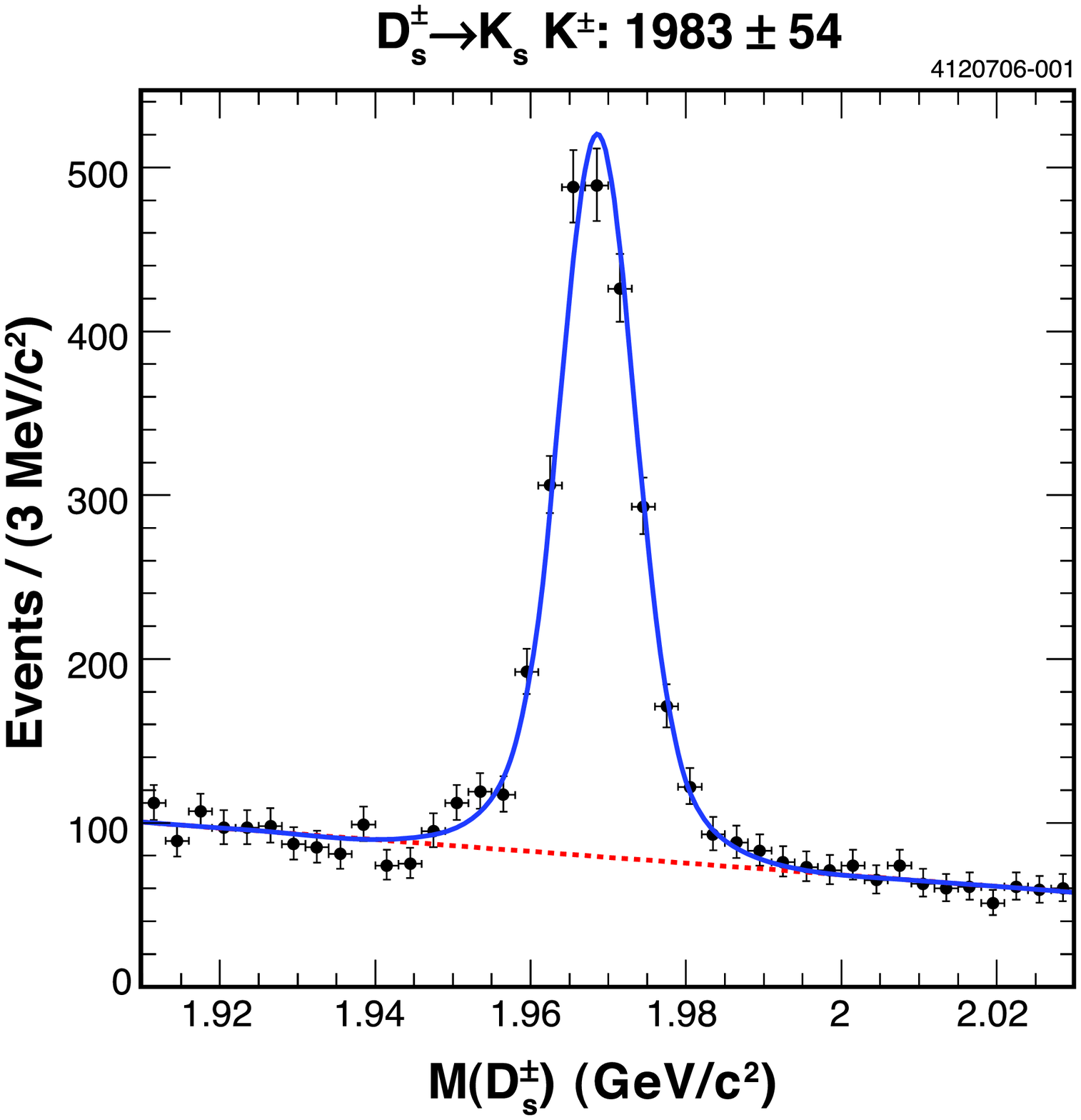} \hfill
  \includegraphics[width=50mm]{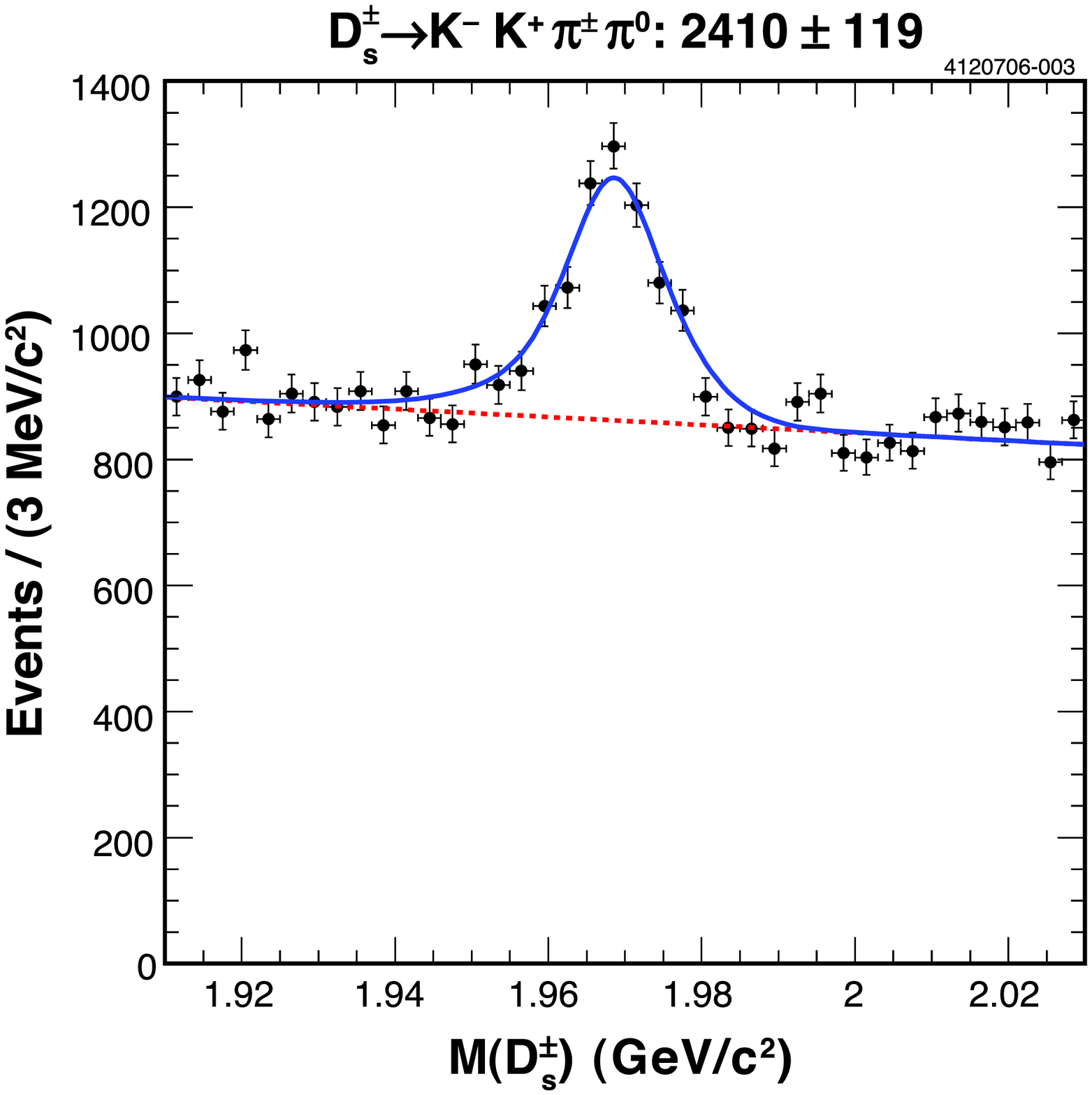} \hfill
  \includegraphics[width=50mm]{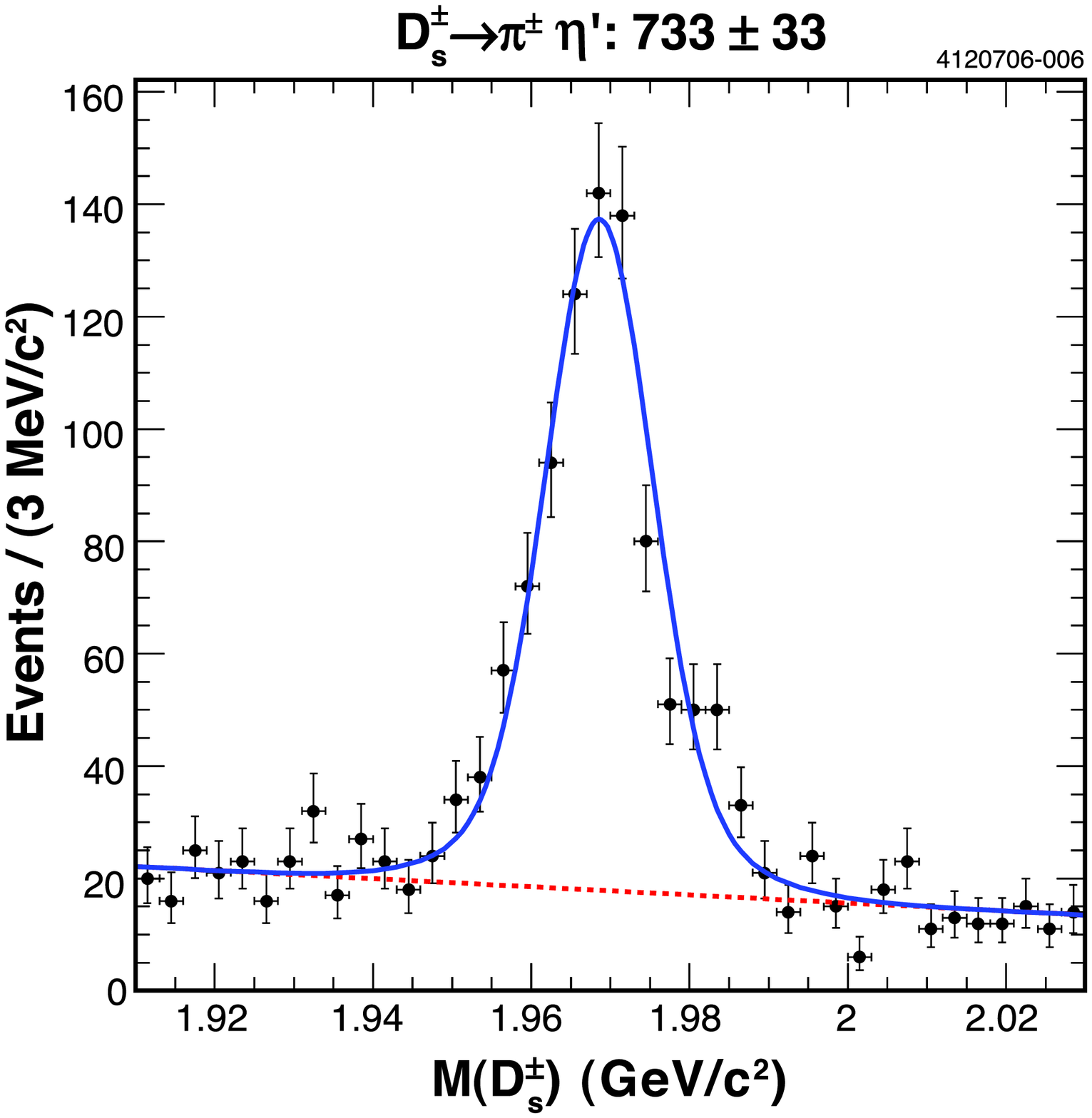} \hfill
  \caption{\label{fig:Ds_HadBF_yields} The three of six invariant mass spectra 
                                       for the ST $D_s^+$ candidates.}
\end{figure}
Good kinematic separation between the $D$ meson pair combinations
can be achieved if we only reconstruct the $D^0$ and $D^+$ candidates and ignore
$\gamma$ and $\pi^0$ from $D^*$ and $D_s^*$ decays. 
It is demonstrated in Fig~\ref{fig:DDmodes_separation}, where
the $K^+K^-\pi^+$ invariant mass is plotted {\it versus} beam constrained mass.
Cutting on invariant mass and $m_{BC}$, Fig.~\ref{fig:Ds_HadBF_mass_spectra}~(left), 
we select events of the process $e^+e^- \to D_s^{\pm}D_s^{*\mp}$. 
In order to measure the $D_s$ meson hadronic branching fractions we use 
the same technique that discussed in Section~\ref{sec:HadBF}.
The number of DT events is counted using 
the signal and background boxes of the 2D invariant mass distribution, 
Fig.~\ref{fig:Ds_HadBF_mass_spectra}~(right).
The number of ST events is extracted from fit to the invariant mass distribution
for each of six modes, as shown in Fig.~\ref{fig:Ds_HadBF_yields} for three of them.
Fit uses a linear shape for background and double Gaussian for signal shape. 
Preliminary results for six $D_s$ modes are shown in Table~\ref{tab:Ds_HadBF} for 195~pb$^{-1}$ sample.
Fair agreement is achieved with results of other experimens~\cite{PDG-2006} for five available modes.

%%%==================================================
\section{Inclusive branching fractions for $\eta$, $\eta '$, and $\phi$} 
\begin{figure}[t]
  \includegraphics[width=160mm]{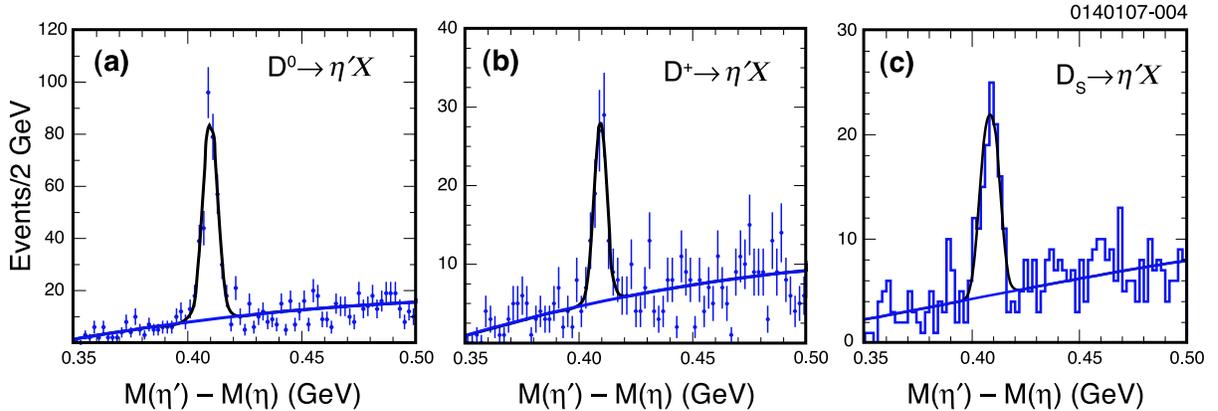}
  \caption{\label{fig:D_InclusiveBF_mass_spectra} The invariant mass spectra 
                                                  for $\eta '$ candidates.}
\end{figure}
\begin{table}[t]
\caption{\label{tab:inclusive_BF} Inclusive branching fractions (in \%) 
                                  of $D$ meson decays and their ratios.}
%\scriptsize
\begin{center}
\begin{tabular}{|c|c|c|c|}
\hline
%\multicolumn{9}{|c|}{ txt } \\
\hline
Mode            & $\eta X$               & $\eta ' X$         &  $\phi X$           \\
\hline
$D^0$           & $9.5\pm0.4\pm0.8$      & $2.48\pm0.17\pm0.21$   & $1.05\pm0.08\pm0.07$\\
$D^+$           & $6.3\pm0.5\pm0.5$      & $1.04\pm0.16\pm0.09$   & $1.03\pm0.10\pm0.07$\\
$D_s^+$         & $23.5\pm3.1\pm2.0$     & $8.7\pm1.9\pm0.8$      & $16.1\pm1.2\pm1.1$  \\
\hline
$D_s^+ / D^0$   & $2.47\pm0.34\pm0.18$   & $3.51\pm0.80\pm0.27$   & $15.3\pm1.6\pm0.8$ \\
$D_s^+ / D^+$   & $3.73\pm0.57\pm0.27$   & $8.37\pm2.23\pm0.64$   & $15.6\pm1.9\pm0.8$  \\
\hline
\end{tabular}
\end{center}
\end{table}
We have measured the branching fractions for the $\eta$, $\eta '$, and $\phi$
mesons inclusive production in $D$-meson decays.
These results are published~\cite{DsHadBF}
for 281~pb$^{-1}$ at $\psi$(3770) and 195~pb$^{-1}$ at 4170~MeV.
We completely reconstruct the tag-side $D$ meson using three decay modes for $D^0$,
five modes for $D^+$, and six modes for $D_s^+$ meson. 
Using the rest particles on the other side, we inclusively reconstruct the 
$\eta \to \gamma \gamma$, $\eta ' \to \pi^+ \pi^- \eta$, and $\phi \to K^+K^-$ decays. 
Invariant masses are used as a signal variables,
as shown for example in Fig.~\ref{fig:D_InclusiveBF_mass_spectra}.
Subtracting background we define relevant yields.
Measured branching fractions are shown in Table~\ref{tab:inclusive_BF}.
We noticed that the $\phi$ meson production branching fractions depend on $\phi$ momentum. 
It is also interesting to note how the decay rate depends on presence of
the $s\bar{s}$ quarks. The $\phi$ meson is almost pure $s\bar{s}$ state, $\eta '$ has a larger
$s\bar{s}$ component than $\eta$. Table~\ref{tab:inclusive_BF} reflects the fact that 
$D_s$ meson prefers to decay in $s\bar{s}$-rich states comparing to the 
$D^0$ and $D^+$ mesons.

%%%==================================================

\section{$D^0$ meson mass measurement}
The precision $D^0$ mass measurement is recently published \cite{D0mass}.
Current mass value~\cite{PDG-2006} has 0.4~MeV/$c^2$ uncertainty, that is a 
result of averaging over several experiments,
dominated by the LGW, MARK II, and NA32 measurements.
In previous experiments the $D^0$ mass was measured using the
$D^0 \to K^-\pi^+$ and $D^0 \to K^-\pi^+\pi^+\pi^-$ decays.
With 281~pb$^{-1}$ sample we use an advantage of the $D^0 \to K^0_S \phi$ 
decay (mass spectrum is shown in Fig.~\ref{fig:D0_mass_spectrum})
which has a small energy released, $M(D^0) - M(\phi) - M(K_S)=347$~MeV/$c^2$, 
that restricts a momentum range of the final particles, 
$(400<P_K,P_\pi <600)$~MeV/$c$.
We estimate a systematic uncertainties in momentum calibration
using inclusively reconstructed  $K^0_S \to \pi^+\pi^-$ in $D$ meson decays.
The maximum uncertainty arises due to the momentum and
$\cos \theta$ dependence for $\pi$-tracks.
The magnetic field calibration is done using invariant mass reconstruction
in the decay $J/\psi \to \mu^+\mu^-$. We also check the $\pi$-meson momentum
calibration using invariant mass reconstruction in the decay
$\psi(2S) \to \pi^+\pi^- J/\psi$.
In both cases we relay on high precision mass measurement~\cite{PDG-2006} 
of the $J/\psi$ and $\psi(2S)$.
We have measured the $D^0$ meson mass value,
$M(D^0) = (1864.847 \pm 0.150_{stat.} \pm 0.095_{syst.} )$~MeV/$c^2$,  
with statistical and systematic uncertainty a few times smaller than the world average.

%%%==================================================

\section{Non-covered results on $D$-meson hadronic decays}
Several topics, listed below, have not been covered in this overview.
Two papers have been recently published,
``Branching fraction for the DCSD  $D^+ \to K^+\pi^-$'' \cite{DptoKpi}, and
``Measurement of interfering $K^{*+}K^-$ and $K^{*-}K^+$ 
amplitudes in the decay $D^0 \to K^+K^-\pi^0$'' \cite{D0toKKpi0}.
The results of several analyses are expected to be presented shortly:
the $D^0$, $D^+$, and $D_s$ hadronic branching fraction measurements for 
single and double Cabibbo suppressed decays;
the quantum correlation analysis of the $D^0\bar{D}^0$ decays;
the Dalitz plot analyses of the decays 
$D^+ \to K^-\pi^+\pi^+$,
$D^+ \to \pi^-\pi^+\pi^+$,
$D_s^+ \to K^-K^+\pi^+$,
$D^0 \to K^0_{(S,L)}\pi^+\pi^-$,
$D^0 \to K^0_S\pi^0\pi^0$,
etc., with and without CP and flavor tags.

%%%==================================================
\section{Summary}

In summary, CLEO-c experiment is taking data from October 2003 until April 2008.
We have collected 4~M of $D\bar{D}$ pairs, 0.3~M of
$D_s^{\pm}D_s^{*\mp}$ pairs at $\sqrt{s}=4170$~MeV, and 28~M of $\psi(2S)$. 
We expect to collect more data in the same energy regions by the end of runs at CESR.
Preliminary results on $D$ meson production and decays are presented.
In most measurements we have reached better precision than the world average.
%%%==================================================

\end{document}